\input psfig.tex
\documentstyle[12pt,aasms4]{article}
\def\simlt{\lower.5ex\hbox{$\; \buildrel < \over \sim \;$}}
\def\simgt{\lower.5ex\hbox{$\; \buildrel > \over \sim \;$}}

\lefthead{Spaans \& van Dishoeck}
\righthead{H$_2$O and O$_2$ in Clumpy Molecular Clouds}

\begin{document}

\title{The Abundance and Emission of H$_2$O and O$_2$ in Clumpy Molecular 
Clouds}

\author{Marco Spaans}

\affil{Kapteyn Institute, P.O. Box 800, 9700 AV Groningen, The Netherlands}

\author{Ewine F.\ van Dishoeck}

\affil{Leiden Observatory, P.O. Box 9513, 2300 RA Leiden, The Netherlands}

\begin{abstract}
Recent observations with the {\it Submillimeter Wave Astronomy
Satellite} indicate abundances of gaseous H$_2$O and O$_2$ in
dense molecular clouds which are significantly lower than found in
standard homogeneous chemistry models.  We present here results for
the thermal and
chemical balance of inhomogeneous molecular clouds
exposed to ultraviolet radiation in which the abundances of H$_2$O and
O$_2$ are computed for various density distributions, radiation field
strengths and geometries.  It is found that an inhomogeneous density
distribution lowers the column densities of H$_2$O and O$_2$ compared
to the homogeneous case by more than an order of
magnitude at the same $A_{\rm V}$. O$_2$ is particularly sensitive to the
penetrating ultraviolet radiation, more so than H$_2$O.
The S~140 and $\rho$ Oph clouds are studied as
relevant test cases of star-forming and quiescent regions. The SWAS
results of S~140 can be accommodated naturally in a clumpy model with
mean density of $2\times 10^3$ cm$^{-3}$ and enhancement $I_{\rm
UV}=140$ compared with the average interstellar radiation field,
in agreement  with
observations of [C I] and $^{(13)}$CO of this cloud.  Additional radiative
transfer computations suggest that this diffuse H$_2$O component is warm,
$\sim 60-90$ K, and can account
for the bulk of the $1_{10}-1_{01}$ line emission observed by SWAS.
The $\rho$~Oph model
yields consistent O$_2$ abundances but too much H$_2$O, even for
[C]/[O]=0.94, if $I_{\rm UV}<10$ respectively $<40$ for a mean density
of $10^3$ respectively $10^4$ cm$^{-3}$. It is concluded that enhanced
photodissociation in clumpy regions can explain the low
H$_2$O and O$_2$ abundances and emissivities found in the large SWAS beam
for extended molecular clouds, but that additional freeze-out of oxygen
onto grains is needed in dense cold cores.
\end{abstract}
{\it subject headings}: ISM: abundances --- ISM: clouds ---
ISM: molecules

\section{Introduction}

In contrast with the major carbon-bearing species (CO, C, C$^+$),
the abundances of the major
oxygen-containing atoms and molecules in general molecular clouds
are still poorly understood.
The best-known species is CO, which reaches large abundances of $\sim
10^{-4}$ when the shielding column of gas exceeds $A_V\approx$1 for a
typical interstellar cloud of $\sim 10^3$ cm$^{-3}$ illuminated by the
average interstellar radiation field, locking up the bulk of the
gas-phase carbon (e.g., Hollenbach \& Tielens 1999, van Dishoeck \& Black
1988).  Theoretical
models indicate that at the edge atomic O is abundant, but that
further into the cloud, more and more of the ultraviolet radiation is
attenuated so that at $A_V\approx 4$ mag, CO ceases to be the dominant
oxygen-bearing species and O$_2$ reaches its peak abundance of
$>10^{-4}$, with H$_2$O present at the level of $\sim 10^{-6}$ (e.g.,
Tielens \& Hollenbach 1985, Lee et al. 1996).
If true, then H$_2$O and O$_2$ could be
major coolants in molecular clouds (Goldsmith \& Langer 1978; Neufeld, Lepp
\& Melnick 1995).

The launch of 
the {\it Submillimeter Wave Astronomy Satellite}
(SWAS) (Melnick et al.\ 2000) has made it
possible to look for emission lines of H$_2$O and O$_2$
in cold molecular clouds. Although the {\it Infrared Space
Observatory} (ISO) observed many H$_2$O lines, it
was sensitive primarily to warm, $\sim 100$ K molecular gas near young stellar
objects where the H$_2$O abundance can be enhanced due to ice
evaporation and high-temperature (shock) reactions (e.g., van Dishoeck
et al.\ 1999, Nisini et al.\ 2000).  Measurements of, or upper limits
on, the abundances of H$_2$O and O$_2$ provide crucial tests of many
aspects of the oxygen chemistry in normal molecular clouds and of the
cloud structure, and hence of our theoretical understanding of such
systems.

The initial SWAS results indicate that the abundances of H$_2$O and
O$_2$ in dense molecular cloud cores are surprisingly low compared
with the above model predictions, about $6\times 10^{-10} - 1\times
10^{-8}$ for H$_2$O and $<$ a few $\times 10^{-7}$ for O$_2$ (Ashby et
al.\ 2000, Snell et
al.\ 2000a, Goldsmith et al.\ 2000). In contrast, ISO observations of
the [O I] 63 and 145 $\mu$m lines suggest substantial abundances of
atomic O in quiescent gas (e.g., Baluteau et al.\ 1997, Caux et al.\
1999).  Bergin et al.\ (2000) propose a model in which freeze-out of
oxygen on dust grains in the form of molecular ices is significant,
with the remaining oxygen in atomic form in the gas, along with CO.
While freeze-out and gas-grain interactions undoubtedly play a role in
dense cold clouds, we present here an alternative explanation based on
clumpy molecular clouds which may be more appropriate for the general
lower density molecular cloud material contained in the large SWAS
beam of $3.3'\times 4.5'$. The inhomogeneous nature of molecular
clouds is well established from observations of extended [C I] and
[C~II] (e.g., Keene et al.\ 1995, Stutzki et al.\ 1988,
Plume et al.\ 1999), and has been modeled
by various groups (e.g., Meixner \& Tielens 1993, Spaans 1996,
St\"orzer et al.\ 1996). It is therefore natural to also explore the
H$_2$O and O$_2$ abundances and emission in such models.

\section{Basic Model Description}

The results presented here were obtained by application of the numerical
code of Spaans (1996), described further in Spaans \& van Dishoeck (1997).
The interested
reader is referred to these papers for a description of the
underlying algorithms.
The code has been specifically designed to solve large
chemical networks with a self-consistent treatment of the thermal balance for
all heating and cooling processes known to be of importance
in the interstellar medium (Spaans \& Ehrenfreund 1999).
The radiative transfer in the cooling lines is solved by means of a Monte
Carlo approach with checks provided by an escape probability method for large
line optical depths. The thermal balance includes photo-electric heating by
dust grains, heating by cosmic rays, absorption of infrared photons by H$_2$O
molecules and their subsequent collisional de-excitation, and gas-grain
heating. The cooling includes the coupling between gas and dust grains
(Hollenbach \& McKee 1989), atomic lines of all metals, molecular lines, and
all major isotopes, as described in Spaans et al.\ (1994) and in Neufeld et al.
(1995) for the regime of very large line optical depth.
All level populations
are computed in statistical equilibrium at a relative accuracy of no less than
$10^{-3}$, an accuracy which is also imposed on the ambient line radiation
field. The combined chemical and thermal balance is required to obey a
convergence criterion between successive iterations of 0.5\%.

The adopted chemical network is based on the UMIST compilation
(Millar, Farquhar \& Willacy 1997), and is well suited for low
temperature ($<$200~K) dense molecular clouds.  A value of 0.3 is
chosen for the branching ratio leading to H$_2$O in the dissociative
recombination of H$_3$O$^+$, with a rate coefficient of $3.3\times
10^{-7}(T/300)^{-0.3}$ cm$^3$ s$^{-1}$. The branching ratio
is consistent with the results of
Vejby-Christensen et al.\ (1997) but higher than the value of 0.05
suggested by Williams et al.\ (1996).
Observations and
(homogeneous) models by Spaans et al.\ (1998) toward the
translucent cloud HD~154368 favor a branching ratio of less than 0.3
(at $3\sigma$) so that the water abundance results presented
here should be regarded as strict upper limits.
The latest
rate coefficients for important neutral-neutral reactions are adopted,
in particular for ${\rm C}+{\rm O}_2\rightarrow {\rm CO}+{\rm O}$ and
${\rm O}+{\rm OH}\rightarrow {\rm O}_2+{\rm H}$ at $2.48\times
10^{-12}(T/300)^{1.54}{\rm e}^{-613/T}$ cm$^3$ s$^{-1}$ and
$1.77\times 10^{-11}{\rm e}^{-178/T}$ cm$^3$ s$^{-1}$,
respectively. The reaction of O$_2$ with neutral sulfur is not found
to be a major sink for molecular oxygen. No gas-grain interactions are
taken into account.
The carbon and oxygen gas-phase fractions are set to $\delta_{\rm
C}=0.33$ and $\delta_{\rm O}=0.35$ (C/O=0.45), but the case of
$\delta_{\rm O}=0.17$ is also considered. These fractions are defined
with respect to the solar values of $[{\rm C}]=4.0\times 10^{-4}$
and $[{\rm O}]=8.3\times 10^{-4}$.  Depletions of other elements are as
in Spaans \& van Dishoeck (1997).

\section{Model Results}

In order to investigate the dependence of O$_2$ and H$_2$O on density,
radiation field and geometry, spherical and slab-like model clouds
were computed for homogeneous and inhomogeneous density distributions
(Spaans 1996). The clumpy models are defined by two parameters: the
volume filling factor $f$, which fixes the fraction of the total
volume that is occupied by the clumps in a two-phase density medium,
and a characteristic clump size $\ell_c$, which fixes the extinction
through an individual clump.  The ratio between the high (h) density
and low (l) density phase is chosen to be $r=20$. It then follows that
the mean total
hydrogen density $n$ obeys $n=fn_{\rm h}+(1-f)n_{\rm l}$.
We model two types of regions: a star-forming cloud where the
radiation field is significantly enhanced due to a nearby O or B star
(e.g., S~140) and a more quiescent region such as $\rho$~Oph.  For the
S~140 extended molecular cloud, Spaans \& van Dishoeck (1997) have
constrained $f<$50\% and $\ell_c>0.2$~pc from observations of [C~I]
and $^{13}$CO (Plume et al.\ 1994), using an average density $n\approx
10^3$ cm$^{-3}$. Recent ISO data of [C~II] and [O~I] confirm
densities of order $10^3$ cm$^{-3}$ in the extended
cloud (Li et al.\ 2001).  We adopt here
$f=30$\% and $\ell_c=0.4$~pc with $n=2\times 10^3$ cm$^{-3}$. 
For models with a larger mean density 
the clump size
$\ell_c$ is decreased by the same factor with which the mean density is
increased. The
enhancement of the radiation field is taken to be $I_{\rm UV}$=140 with
respect to the Draine (1978) field, consistent with the enhancement
at the edge of the cloud near the B0V star HD~211880. This enhancement factor
decreases to $I_{\rm UV}\approx 30-50$ further
into the extended cloud due to geometric dilution.
The average interstellar radiation field incident from other directions
has been included as well.  The $\rho$ Oph
cloud has been studied by several groups (see Liseau et al.\ 1999) and
is modeled with $f=20$\% and $r=30$, more clumpy than S~140 but with
an identical value for $\ell_c$. We adopt $I_{\rm UV}\approx 10$ and
$n=10^4$ cm$^{-3}$. It thus provides an interesting comparison.

Figures 1--2 present the basic results of this work. For the
inhomogeneous models the shell/slice averaged abundances are shown for
the sphere/slab as functions of $A_V$.
It was found that the
dispersion around the mean values by taking different cuts is not
larger than about 50\%, so that the systematics discussed below are robust.
The following trends can be identified. First, for the same
$I_{\rm UV}$ and $n$, a strong decrease in the abundances of O$_2$ and
H$_2$O is found in the inhomogeneous models, although less so for
H$_2$O.  The reason is that the destruction of O$_2$ remains dominated
by photodissociation to large extinctions and that the clumpy
structure allows ultraviolet photons to penetrate to larger
depths. The effect spans a factor of 30 or more.
Second, the central abundances of H$_2$O and O$_2$ are decreased by a
factor of 2-4 as the geometry is changed from a plane-parallel slab to
a sphere. The reason is simply that the shielding
column is larger for some rays emanating from the center of a slab compared
to a sphere.
A clumpy density distribution is quite effective in redistributing the
directed ultraviolet radiation field from a (point-like) source into a
more isotropic distribution at a depth of a few magnitudes of visual
extinction (Spaans 1996).
Third, a systematic trend is observed with varying dissociation
parameter $U=I_{\rm UV}/n$. When this number is small, $<5\times 10^{-3}$,
i.e., at low $I_{\rm UV}$ and/or high $n$,
H$_2$O and O$_2$ rise sharply at the edge of the cloud, and
H$_2$O quickly reaches the regime where its removal is dominated by
chemistry rather than photodissociation.  The total observed column of
H$_2$O and O$_2$ varies greatly with total extinction along the line
of sight and it would be difficult to distinguish between homogeneous
large $U$ and inhomogeneous small $U$ situations based on H$_2$O and
O$_2$ alone. Of course, large $U$ regions are bright in lines such as
[C~II] 158 $\mu$m, [O~I] 63 and 145 $\mu$m and high-$J$ CO, whereas
small $U$ regions tend to have
more prominent [C~I] 609 $\mu$m emission, in particular when they are
clumpy.

\section{Discussion and Comparison with Observations}

\subsection{H$_2$O and O$_2$ abundances}

The observational results represent abundances integrated
over depth. Thus, mass-weighted abundances for the spherical models have
been computed for comparison with observations. 
Specifically, the mean abundance $X(r)$ enclosed within
some fractional radius $r$ is given by
integration of the abundance $A(r)$ through
$X(r)$=$\int_0^r A(r')r'^2dr'$/$\int_0^r r'^2dr'$, 
where 0 is at the center and 1 at the edge of the model cloud.
Note that only $1/8$
of the mass is contained inside the half radius of the spherical
constant density model cloud, while this would be $1/2$ for a $1/r^2$
density profile.

For the clumpy, spherical model of Figure 1 (top and middle) the above
approach leads to a predicted mean H$_2$O abundance
$X(1)=3\times 10^{-8}$ and
an O$_2$ abundance of $4\times 10^{-10}$ for a typical total depth
of $A_V=10$~mag. Snell et al.\ (2000a)
detected H$_2$O emission from the S~140 dense star-forming core and
PDR region with SWAS, but not from the surrounding extended
molecular cloud.  Their inferred abundances of $\sim 10^{-8}$ (see also the
detailed analysis by Ashby et al.\ 2000) for the
core and $< 10^{-8}$ for the extended cloud are only a factor of three
lower than our model results and consistent within the factor of three or
more uncertainty in their assumed density (the inferred H$_2$O
abundance scales with the inverse of density).
Figure 1c (bottom) shows the effect of increasing the
density to $5\times 10^4$ cm$^{-3}$ for the S~140 environment.  The
smaller value of $U$ causes photodissociation to play a lesser role
in the removal of O$_2$ and H$_2$O, and hence clumpiness is of less
importance. The model (mass weighted) abundances are $3\times 10^{-7}$ and
$4\times 10^{-8}$ for H$_2$O and O$_2$, respectively.

The model for $\rho$ Oph appears less successful in
reproducing the observed abundances of O$_2$ and H$_2$O. The mass
weighted H$_2$O abundance for Figure 2c (bottom) is $3\times
10^{-7}$, a factor of $\sim$100 above the measured value.  Panels 2a and b
(top and middle), which are representative of larger values of
$U$ and roughly consistent with the results of Liseau et al.\ (1999) who found
$n_{\rm H}=3\times 10^3-3\times 10^4$ cm$^{-3}$,
yield mass-weighted H$_2$O abundances of $5\times 10^{-8}$ and $2\times
10^{-8}$, respectively.
From the discussion in \S 3, it follows that the 
S~140 case yields better agreement
with the observations mainly because of its larger illumination (and $U$
parameter). The clumpy models enhance the importance of photodissociation
of H$_2$O and O$_2$ deeper into the cloud,
and therefore work best for a strong incident radiation field.

In all inhomogeneous models with $n\approx 10^3$ cm$^{-3}$, the mass weighted
abundance of O$_2$ is below $10^{-7}$,
both for S~140 and $\rho$ Oph,
consistent with the overall SWAS upper limits as well as the specific
upper bounds for S~140, $<7\times 10^{-7}$, and $\rho$ Oph, $<3\times
10^{-7}$ (Goldsmith et al.\ 2000). Only the quiescent model of Figure
2c (bottom) with high density and low $I_{\rm UV}$, yields a mass
weighted O$_2$ abundance of $10^{-6}$. This is a factor of
a few above the observational limit (Goldsmith et al.\ 2000), but this
is easily alleviated by a modest increase in the [C]/[O] ratio (Bergin et
al.\ 2000). To test the latter possibility,
we have run a model with an oxygen depletion of 0.17 and
[C]/[O]=0.94. A decrease in the mass weighted H$_2$O (O$_2$) abundance of a
factor of 5 (20) is found,
bringing the model in agreement with the observations for S~140 and
factors of 20, 4 and 2 too high for H$_2$O toward $\rho$ Oph
in models 2a, b and c, respectively, although a smaller value for the
H$_3$O$^+$ branching ratio of 0.05 instead of 0.3
could further reduce these discrepancies.

\subsection{H$_2$O and O$_2$ emission}

In order to verify whether the low abundances of H$_2$O and O$_2$ found for
the clumpy model clouds can also reproduce the bulk of the observed  emission,
radiative transfer computations have been performed using a Monte Carlo method
(Spaans 1996). It was found that two effects dominate the emissivity of the
H$_2$O $1_{10}-1_{01}$  line: temperature and geometry. The temperature is
very important because it influences both the ortho-para ratio of the main
collision partner H$_2$, 0.1 at 30 K and 1.0 at 80 K (Neufeld \& Sternberg
1999), as well as the collisional excitation rate. The H$_2$ ortho-para ratio
is a crucial parameter to compute rigorously from the thermal balance and FUV
pumping since the o-H$_2$ $J$=1 collision rates with H$_2$O are more than an
order of magnitude larger than the p-H$_2$ $J$=0 rates (Phillips, Maluendes \&
Green 1996). The geometry is important because the $1_{10}-1_{01}$ line is
optically thick but effectively thin. This implies that
line photons emitted by the warm edges of a spherical clump can interact with
water molecules in the colder interior, and so constitute an excitation term
that raises the level population in the $1_{10}$ state.
Note in this that the
curvature of the sphere causes the warm gas to cover $4\pi$ steridians, unlike
the slab case.
It is found that a spherical geometry, for the S~140 temperature
profile given below, raises the $1_{10}-1_{01}$ line strength with a factor of
2.4 compared to that from the plane-parallel slab.
Furthermore,
for identical physical conditions, a plane-parallel slab model yields
a temperature profile that is systematically 20-30\% cooler than for the
corresponding spherical case, mainly because the latter possesses more lower
extinction lines of sight.

The thermal balance yields a temperature of 150 (80) K at the edge of S~140
($\rho$~Oph) and 20 (10) K at its center. The mass weighted H$_2$O temperatures
are $\sim 80$ K and $\sim 30$ K for S~140 and $\rho$~Oph, respectively. For
S~140, these values refer to the model of Fig.\ $1a/b$, and for $\rho$~Oph to
the model of Fig.\ $2b$. The S~140 value is significantly larger than the
typical temperature in dense cores, $\sim$ 30~K. The resulting integrated line
intensities convolved with the SWAS beam are 1.6 K km s$^{-1}$ and 3.8 K km
s$^{-1}$ for S~140 and $\rho$~Oph, respectively. The clumpy S~140 model can
thus reproduce the observed value (1.7 K km s$^{-1}$) quite nicely, but the
clumpy $\rho$~Oph H$_2$O emission, just as the abundance, is too large compared
to the observed value of 0.8 K km s$^{-1}$. Use of a plane-parallel geometry
would reduce the discrepancy for $\rho$ Oph significantly. In both instances,
the bulk of the signal in the SWAS beam is associated with the PDR.
In fact, {\it the H$_2$O emission arises mainly from the warm, $\sim$60-90 K,
edges of irradiated clumps}.
Therefore, a strong correlation should exist between the SWAS H$_2$O
emission and high$-J$ (e.g.\ 5-4, 6-5) CO emission. Such a correlation is in
fact observed toward M~17SW (Snell et al.\ 2000b).
It is the elevated temperatures at the surfaces of the
clumps that makes the results so sensitive to the ortho-para ratio.
For the case of S~140, the extended molecular cloud gives an emission that is a
factor of 30 weaker at a position $4'$ north-east of the PDR, due to the lower
ambient temperature, consistent with observations.
The O$_2$ $N_J=3_3-1_2$ line is truly optically thin and yields, due to its
simpler excitation, more robust results.
We find $1.8\times 10^{-5}$ K km s$^{-1}$ and $2.3\times 10^{-3}$ K km
s$^{-1}$ for the S~140 and $\rho$~Oph case, respectively.


ISO H$_2$O observations suggest that much of H$_2$O
toward star-forming cores such as S~140
is associated with a small region with enhanced
H$_2$O surrounding the young stellar object(s)
(e.g., Wright et al.\ 1997, Ceccarelli et al.\ 1999). Since this emission is
subsequently beam-diluted in the large SWAS beam, 
its contribution could be small and still
be consistent with the fact that the lower density gas in the SWAS beam is
capable of reproducing the bulk of the detected water emission by SWAS.
We have verified this by adding a clumpy star-forming core with $n=10^4$
cm$^{-3}$ and a size of $\sim 0.5$ pc (Zhou et al.\ 1994), and the same values
for $f$ and $r$, to the S~140 extended
molecular cloud as in Spaans \& van Dishoeck (1997).
Even though the mean water abundance goes up by a factor of 30, the
lower kinetic temperature and smaller spatial extent yield a relative
contribution of $\sim 20$\% after convolution with the SWAS beam. Furthermore,
the Zhou et al.\ CS observations can be well reproduced by this clumpy core
that includes densities of up to $\sim 10^5$ cm$^{-3}$ and temperatures that
are still $\sim 30-40$ K.
\footnote{At even higher densities ($\sim 10^7$ cm$^{-3}$) and smaller scales
($\sim 200-400$ AU), the H$_2$O abundance becomes even larger and the region
emits thermally at about 100 K.
Strong beam-dilution limits this
contribution to the SWAS signal to less than 2\%  at 
the distance of $\sim 910$ pc for S~140.}


Bergin et al.\ (2000) have argued against the importance of
photodissociation because it would affect the abundances of other
species such as NH$_3$ as well.  The S~140 model, including the
star-forming core as above, was checked for
consistency with the observational NH$_3$ values of $2-9\times 10^{14}$
cm$^{-2}$ (Ungerechts, Winnewisser \& Walmsley 1986). Our model value of
$\sim 4\times 10^{14}$ cm$^{-2}$ is not in conflict with observations.
In addition, even a moderate enhancement in NH$_3$ formation
due to grain-surface chemistry is likely to
resolve any mild discrepancy.  For the extended molecular cloud
outside the core there are no observational constraints for ammonia.
Future observational and theoretical studies
should investigate other species whose abundances are sensitive to
photodissociation, such as the CN/HCN and C$_2$H/HCO$^+$ ratios (e.g.,
Jansen et al.\ 1995). 
This work has not addressed the effects of time-dependent chemistry and
freeze-out on the O$_2$ and H$_2$O abundances (Bergin et al.\ 2000, Viti 2000,
private communication). Our results suggest that clumpiness is a viable
alternative explanation for extended lower-density molecular clouds in
star-forming regions like S~140 with high
radiation field, but they also strengthen the case
for time-dependent chemistry and
freeze-out for a more quiescent dense region like $\rho$ Oph.
The freeze-out models may in fact benefit from adoption of a clumpy
structure since the central dust temperature is enhanced by the higher
internal ultraviolet field, thus providing a mechanism to moderate the
very rapid freeze-out.
Future observatories such as ODIN and the HIFI instrument on the
{\it Far Infrared and Submillimeter Telescope} (FIRST) will have
increased sensitivity, and, in the case of FIRST, a much smaller beam. 
Therefore, more definite tests of the various classes of models will be
forthcoming in this decennium.

\acknowledgments
We are grateful to Ted Bergin, Gary Melnick, Matt Ashby and Serena Viti for
stimulating discussions regarding the SWAS results, to Steven Doty for
assistance with the chemistry, and to the referee, Ted Bergin, for his
constructive comments that helped to improve this paper.

\newpage

\begin{figure}
\label{figure1}
\caption{Models appropriate of the S~140 extended molecular cloud.
1a (top): Model cloud with $n=2\times 10^3$ cm$^{-3}$ and $I_{\rm UV}=140$.
Solid lines are clumpy spherical model, dashed lines 
homogeneous spherical
model; 1b (middle): Model cloud with $n=2\times 10^3$
cm$^{-3}$ and $I_{\rm UV}=140$.  Solid lines are clumpy spherical model,
dashed lines clumpy slab model; 1c (bottom): Model cloud with
$n=5\times 10^4$ cm$^{-3}$ and $I_{\rm UV}=140$.  Solid lines are clumpy
spherical model, dashed lines homogeneous slab model.}
\end{figure}

\begin{figure}
\label{figure2}
\caption{Models appropriate for the $\rho$~Oph
cloud. 2a (top): cloud with $n=10^3$ cm$^{-3}$ and $I_{\rm UV}=10$.
Solid lines are clumpy spherical model, dashed lines homogeneous slab
model; 2b (middle): Model cloud with $n=10^4$ cm$^{-3}$ and $I_{\rm
UV}=40$.  Solid lines are clumpy spherical model, dashed lines homogeneous
spherical model;  2c (bottom): Model cloud with $n=10^4$
cm$^{-3}$ and $I_{\rm UV}=10$.  Solid lines are clumpy spherical model,
dashed lines homogeneous spherical model.}
\end{figure}

\clearpage
\centerline{\psfig{figure=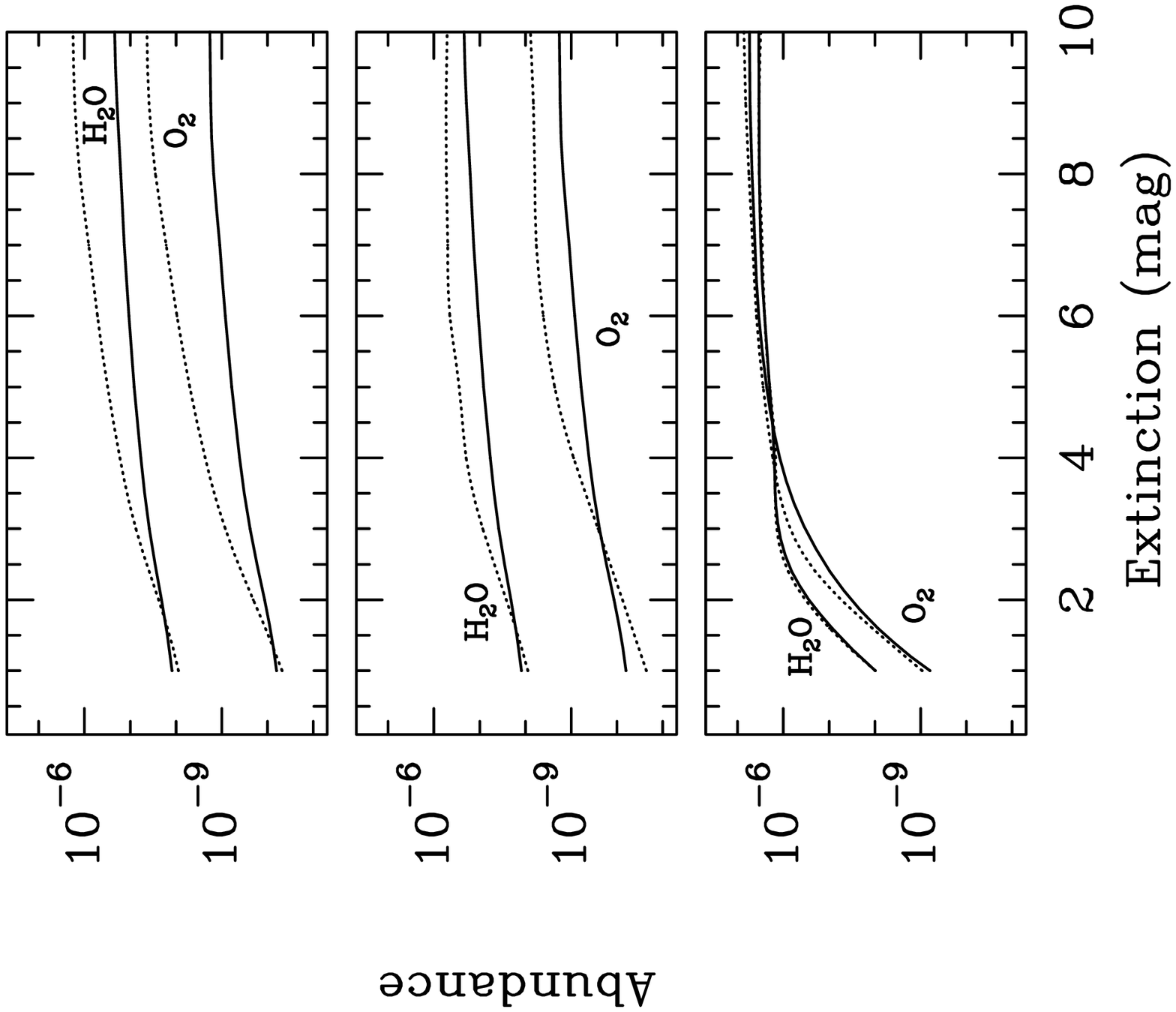,height=18.0truecm,angle=-90}}

\clearpage
\centerline{\psfig{figure=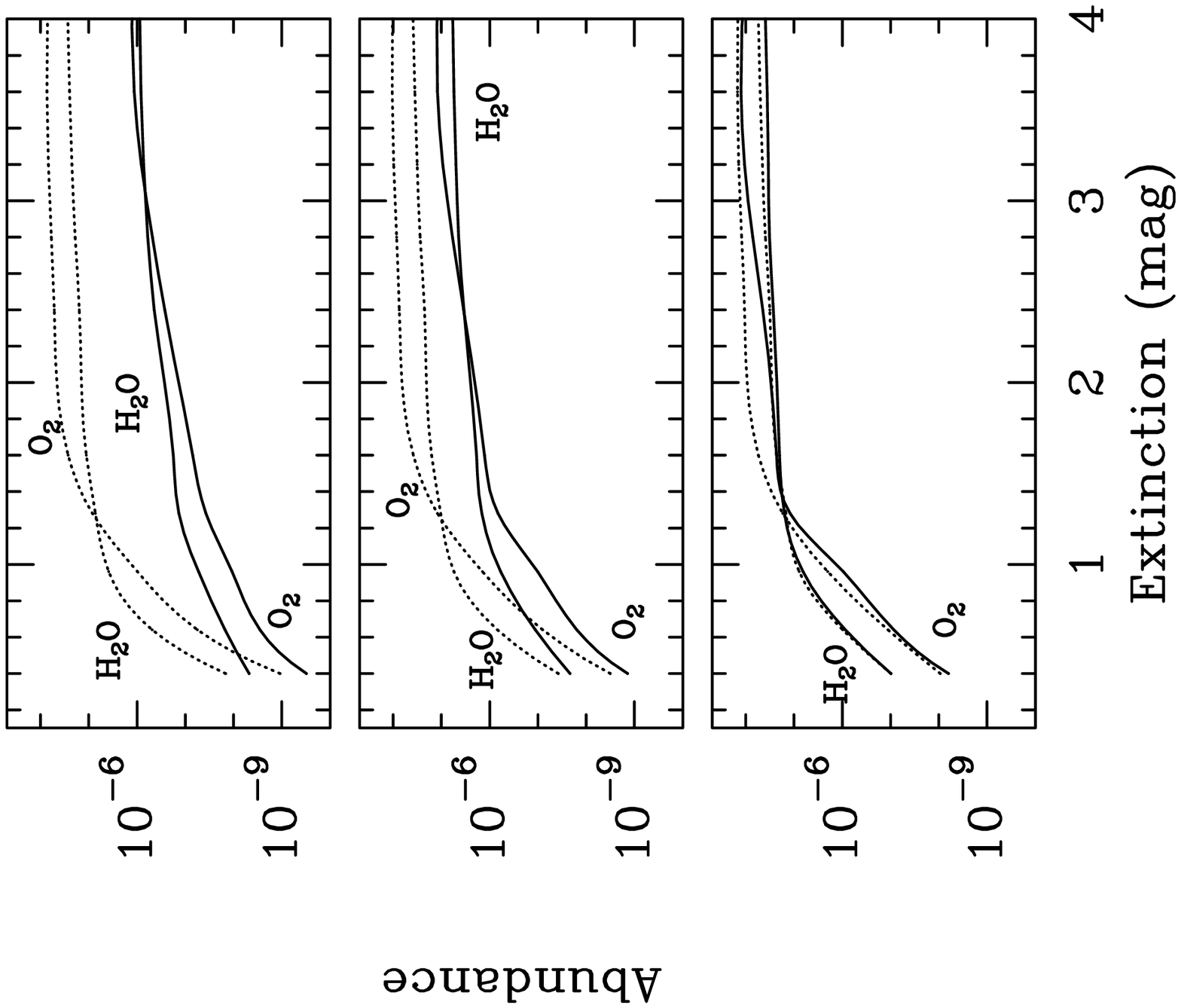,height=18.0truecm,angle=-90}}

\end{document}